\documentclass{rjparticle}
\usepackage{graphicx}

\usepackage{epstopdf}

\newtheorem{Theorem}{Theorem}[section]
\newtheorem{Lemma}[Theorem]{Lemma}
\newtheorem{Proposition}[Theorem]{Proposition}
\newtheorem{Definition}[Theorem]{Definition}

\newcommand{\bDelta}{\boldsymbol{\Delta}}
\newcommand{\im}{\mathrm{im}}

\newcommand{\bcE}{\boldsymbol{\check{E}}}

\newcommand{\bp}{\begin{Proposition}}
\newcommand{\ep}{\end{Proposition}}
\newcommand{\bl}{\begin{Lemma}}
\newcommand{\el}{\end{Lemma}}
\newcommand{\bt}{\begin{Theorem}}
\newcommand{\et}{\end{Theorem}}
\newcommand{\bd}{\begin{Definition}}
\newcommand{\ed}{\end{Definition}}

\newcommand{\eqdef}{\stackrel{{\rm def.}}{=}}

\DeclareFontFamily{U}{rsf}{}
\DeclareFontShape{U}{rsf}{m}{n}{<5> <6> rsfs5 <7> <8> <9> rsfs7 <10-> rsfs10}{}
\DeclareMathAlphabet\Scr{U}{rsf}{m}{n}

\def\cU{\mathcal{U}}
\def\cW{\mathcal{W}}

\def\R{\mathbb{R}}

\def\rk{{\rm rk}}

\def\dd{\mathrm{d}}

\def\AdS{\mathrm{AdS}}

\def\Int{\mathrm{Int}}
\def\Fr{\mathrm{Fr}}
\def \fr{\mathrm{fr}}

\def\Sing{\mathrm{Sing}}

\def\ind{\mathrm{ind}}
\def\bcF{\bar{\cF}}

\newcommand{\nn}{\nonumber}

\newcommand{\beqa}{\begin{eqnarray*}}
\newcommand{\eeqa}{\end{eqnarray*}}
\newcommand{\beqan}{\begin{eqnarray}}
\newcommand{\eeqan}{\end{eqnarray}}

\def\per{\mathrm{per}}

\def\cC{{\mathcal C}}

\def\Spin{\mathrm{Spin}}

\def\cD{\mathcal{D}}

\def\cE{\mathcal{E}}

\def\cN{\mathcal{N}}

\def\cF{\mathcal{F}}
\def\cC{\mathcal{C}}

\def\G_2{\mathrm{G_2}}

\def\cL{\mathcal{L}}
\def\cS{\mathcal{S}}

\def\mF{\mathbf{F}}
\def\momega{{\boldsymbol{\omega}}}

\def\f{\mathfrak{f}}
\def\m{\mathrm{min}}
\def\M{\mathrm{max}}

\def\dotC{\overset{\bullet}{C}}

\title{Foliated backgrounds for M-theory compactifications (II)}
\author[1,2]{E. M. Babalic}
\author[3]{C. I. Lazaroiu}
\affil[1]{Department of Theoretical Physics,\\
``Horia Hulubei'' National Institute for Physics and Nuclear Engineering,\\
Reactorului 30, POB-MG6, Magurele-Bucharest 077125, Romania}
\affil[2]{Department of Physics, University of Craiova, 13 Al. I. Cuza Str., Craiova 200585, Romania\\
{\em Email}: mbabalic@dthph.nipne.ro}
\affil[3]{Center for Geometry and Physics,Institute for Basic Science (IBS)\\Pohang 790-784, Republic of Korea\\
{\em Email}: calin@ibs.re.kr}

\begin{document}
\maketitle
\begin{abstract}
We summarize the foliation approach to $\cN=1$ compactifications of
eleven-dimensional supergravity on eight-manifolds $M$ down to
$\AdS_3$ spaces for the case when the internal part $\xi$ of the
supersymmetry generator is chiral on some proper subset $\cW$ of
$M$. In this case, a topological no-go theorem implies that the
complement $M\setminus \cW$ must be a {\em dense} open subset, while
$M$ admits a {\em singular} foliation ${\bar \cF}$ (in the sense of
Haefliger) which is defined by a closed one-form $\momega$ and is
endowed with a longitudinal $G_2$ structure. The geometry of this
foliation is determined by the supersymmetry conditions. We also
describe the topology of $\bcF$ in the case when $\momega$ is a Morse
form.
\end{abstract}

\section{Introduction}
We describe the extension of the results of \cite{g2} (which were
summarized in \cite{TIM14}) to the general case when the internal part
$\xi$ of the supersymmetry generator is allowed to become chiral on
some locus $\cW\subset M$. Assuming that $\cW\neq M$, i.e. that $\xi$
is not everywhere chiral, we showed in \cite{g2s} that, at the
classical level, the Einstein equations imply that the chiral locus
$\cW$ must be a set with empty interior, which therefore is negligible
with respect to the Lebesgue measure of the internal space. As a
consequence, the behavior of geometric data along this locus can be
obtained from the non-chiral locus $\cU\eqdef M\setminus \cW$ through
a limiting process. The geometric information along the non-chiral
locus is encoded \cite{g2} by a regular foliation $\cF$ which carries
a longitudinal $G_2$ structure and whose geometry is determined by the
supersymmetry conditions in terms of the supergravity four-form field
strength.  When $\emptyset \neq \cW\subsetneq M$, $\cF$ extends to a
singular foliation $\bcF$ of the whole manifold $M$ by adding leaves
which are allowed to have singularities at points belonging to the
chiral locus $\cW$. This singular foliation ``integrates'' a
cosmooth\footnote{Note that $\cD$ is {\em not} a singular distribution
  in the sense of Stefan-Sussmann (it is cosmooth rather than
  smooth). } singular distribution $\cD$ (a.k.a. generalized
sub-bundle of $TM$), defined as the kernel distribution of a closed
one-form $\momega$ which belongs to a cohomology class $\f\in
H^1(M,\R)$ determined by the supergravity four-form field
strength. The vanishing locus $\Sing(\momega)$ of $\momega$ coincides
with the chiral locus $\cW$. In the most general case, $\bcF$ can be
viewed as a Haefliger structure \cite{Haefliger} on $M$. The singular
foliation $\bcF$ carries a longitudinal $G_2$ structure, which is
allowed to degenerate at the singular points of singular leaves. On
the non-chiral locus $\cU$, the problem can be studied using the
approach of \cite{g2} or the approach advocated in \cite{Tsimpis},
which makes use of two $\Spin(7)_\pm$ structures.  The results of
\cite{g2} agree with those of \cite{Tsimpis} along this locus, as shown 
in \cite{g2s} by direct computation, upon using a certain
``refined parameterization'' of the flux components.

The topology of singular foliations defined by a closed one-form can
be extremely complicated in general, but it is better understood when
$\momega$ is a Morse one-form. In the Morse case, the singular foliation
$\bcF$ can be described using the {\em foliation graph}
\cite{MelnikovaThesis} associated to a certain decomposition
of $M$, which provides a combinatorial way to encode some important
aspects of the foliation's topology --- up to neglecting the
information contained in the so-called {\em minimal components} of the
decomposition, components which are expected to possess an as yet
unexplored non-commutative geometric description.

We work in the same compactification set-up as in \cite{Tsimpis,
  MartelliSparks}, with the same notations and conventions as in
\cite{g2,g2s}.  In such warped flux compactification, the
supersymmetry conditions are equivalent with certain algebraic and
differential constraints on some differential forms constructed from $\xi$.

\section{Globally valid parameterization of a Majorana spinor on $M$}
\label{sec:fierz}

Fixing a Majorana spinor $\xi\in \Gamma(M,S)$ which is everywhere of
norm one, we consider the inhomogeneus differential form (see
\cite{g2s}):
\beq
\label{checkE}
\check{E}_{\xi,\xi} \eqdef \check{E}=\frac{1}{16} \sum_{k=0}^8 \bcE^{(k)}~~\nn
\eeq
where we use the following notations for the non-vanishing rank components : 
\beq
\label{forms8}
\bcE^{(0)}=||\xi||^2=1~~,~~ \bcE^{(1)} \eqdef V ~~,~~ \bcE^{(4)}\eqdef Y ~~,
~~ \bcE^{(5)}\eqdef Z~~,~~  \bcE^{(8)}\eqdef b\nu~~,\nn
\eeq
with $b$ a smooth real valued function defined on $M$

The Fierz identities can be fully described \cite{rec} by the conditions:
\beq
\label{Esquare}
\check{E}^2=\check{E}~~,~~~\cS(\check{E})=1~~,~~\tau(\check{E})=\check{E}~~
\eeq
and can be shown to be equivalent with a set of relations which hold
globally on $M$ and which differ from those obtained in \cite{g2} by
including the limit when $b=\pm 1$, equivalently when $V=0$.  In particular, the
geometric data along $\cW$ can be recovered from $\cU$ through a
limiting process which is described in \cite{g2s}. We will not give
any details of the calculations here (see \cite{g2,g2s}), but will focus on
describing the geometric meaning of the results.

\paragraph{The chirality decomposition of $M$.} 
Let $S^\pm\subset S$ be the positive and negative chirality  subbundles of $S$, which give the 
orthogonal decomposition $S=S^+\oplus S^-$.  Decomposing a
normalized spinor as $\xi=\xi^++\xi^-$ with $\xi^\pm\in \Gamma(M,S^\pm)$, we have:
\beqn
||\xi||^2=||\xi^+||^2+||\xi^-||^2=1~~,~~b=||\xi^+||^2-||\xi^-||^2~~,
\eeqn
which give: 
\beq
\label{xipmnorms}
||\xi^\pm||^2=\frac{1}{2}(1\pm b)~~.
\eeq
Notice that $b$ equals $\pm 1$ at a point $p\in M$ iff $\xi_p\in
S^\pm_p$.  Since $||V||^2=1-b^2$ (implied by \eqref{Esquare}), the one-form $V$ vanishes at $p$
iff $\xi_p$ is chiral i.e. iff $\xi_p\in S_p^+\cup S_p^-$. Consider
the {\em non-chiral locus} (an open subset of $M$):
\beqn
\cU \eqdef \{p\in M|\xi^+_p\neq 0~\mathrm{and}~~\xi_p^-\neq 0\}=\{p\in M|V_p\neq  0\}=\{p\in M||b(p)| < 1\}~~,
\eeqn
and its closed complement, the {\em chiral locus}:
\beqn
\cW\eqdef M\setminus \cU=\{p\in M|\xi^+_p=0~\mathrm{or}~\xi^-_p=0\}=\{p\in M|V_p=0\}=\{p\in M| |b(p)|=1\}~~.
\eeqn
The chiral locus $\cW$ decomposes further as a disjoint union of two closed
subsets, the {\em positive and negative chirality loci} 
$\cW=\cW^+\sqcup \cW^-$, where:
\beqa
\cW^\pm\eqdef \{p\in M|\xi_p\in S^\pm_p\}=\{p\in M| b(p)=\pm 1\}=\{p\in M|\xi_p^\mp=0\}~~.
\eeqa
Since $\xi$ does not vanish on $M$, we have: 
\beqn
\cU^\pm\eqdef \cU\cup \cW^\pm=\{p\in M|\xi^\pm_p\neq 0\}~~.
\eeqn

\section{A topological no-go theorem}

We remind the reader the warped compactification ansatz for the field strength  $\mathbf{G}$ of the supergravity 3-form field:
\beq
\label{Gansatz}
\mathbf{ G} = \nu_3\wedge \mathbf{f}+\mF~~,~~~~\mathrm{with}~~ 
\mF\eqdef e^{3\Delta}F~~,~~\mathbf{f}\eqdef e^{3\Delta} f~~
\eeq
where $\nu_3$ is the volume form of the $\AdS_3$ space,  $f\in \Omega^1(M)$, $F\in \Omega^4(M)$, while 
$\Delta$ is the warp factor. Another quantity that appears in the relations is 
$\kappa$, a positive real parameter describing the $\AdS_3$ space ($\kappa$ becomes zero in the Minkowski limit).

\paragraph{Theorem \cite{g2s}.} Assume that the supersymmetry conditions, the Bianchi identity and
equations of motion for $G$ as well as the Einstein equations are
satisfied. Then there exist only the following possibilities:
\begin{enumerate}
\item The set $\cW^+$ coincides with $M$ and hence $\cW^-$ and $\cU$
  are empty. In this case, $\xi$ is a chiral spinor of positive
  chirality which is covariantly constant on $M$ and we have
  $\kappa=f=F=0$ while $\Delta$ is constant on $M$.
\item The set $\cW^-$ coincides with $M$ and hence $\cW^+$ and $\cU$
  are empty. In this case, $\xi$ is a chiral spinor of negative
  chirality which is covariantly constant on $M$ and we have
  $\kappa=f=F=0$ while $\Delta$ is constant on $M$.
\item The set $\cU$ coincides with $M$ and hence $\cW^+$ and $\cW^-$
  are empty.
\item At least one of the sets $\cW^+$ and $\cW^-$ is non-empty but
  both of these sets have empty interior. In this case, $\cU$ is dense
  in $M$ and the union $\cW=\cW^+\cup\cW^-$ coincides with the
  topological frontier $\Fr(\cU)=\fr(\cU)={\bar \cU}\setminus \cU$ ~of ~$\cU$.
 \end{enumerate} 
 
\noindent The proof relies on the analysis of the supersymmetry
conditions (see \cite{g2s}).  Cases 1 and 2 correspond to the
classical limit (the limit when the quantum correction required by
M5-brane anomaly cancellation is negligible) of the well-known
compactifications of \cite{Becker1}. Case 3 was studied in \cite{g2}
(having been pioneered in \cite{MartelliSparks} -- where, however, a
complete solution was not given). In this paper we concentrate
on Case 4 (which was first considered in \cite{Tsimpis}, though from a different perspective).
Hence, from now on we assume:
\beqn
M={\bar \cU}=\cU\sqcup \cW~~,~~\cW=\Fr \cU~~.
\eeqn
The foliation approach to this case (which we describe below) gives a
handle on both local and {\em global} aspects of the geometry of $M$
and shows that, due to global aspects, the relation between such
compactifications and 7-dimensional compactifications of M-theory is
much more subtle than one may imagine at fist sight.

\section{The singular foliation $\bcF$}

The one-form $V$ determines a generalized distribution $\cD$ (generalized sub-bundle of $TM$) which is defined through:
\beq
\cD_p\eqdef{\ker V_p}~~,~~\forall p\in M~~.
\eeq
This singular distribution is {\em cosmooth} (rather than smooth) in
the sense of \cite{Drager}. The set of regular points of $\cD$ equals the
non-chiral locus $\cU$ and we have:
\begin{align}
\rk\cD_p&=7~~\mathrm{when}~~p\in\cU~~,\nn\\
\rk\cD_p&=8~~\mathrm{when}~~p\in \cW~~.\nn
\end{align}
In particular, the restriction $\cD|_\cU$ is a regular Frobenius
distribution.  As in \cite{g2}, we endow
$\cD|_\cU$ with the orientation induced by that of $M$.
One can show that the one-form: 
\beq
\momega\eqdef 4\kappa e^{3\Delta} V~~
\eeq
satisfies the following relations, which hold globally on $M$ as a
consequence of the supersymmetry conditions:
\beq
\label{meq}
\dd\momega=0~~,~~\momega~=\mathbf{f}-\dd\mathbf{b}~~,~~\mathrm{where}~~\mathbf{b}\eqdef e^{3\Delta}b~~.
\eeq
As a result of the first equation, the generalized distribution
$\cD=\ker V=\ker\momega$ determines a singular foliation $\bcF$ of
$M$, which degenerates along the chiral locus $\cW$; since $\cD$ is
cosmooth rather than smooth, this singular foliation can be described
as a Haefliger structure \cite{Haefliger} (see \cite{g2s} for
details), thus it is not a singular foliation in the sense of
Stefan-Sussmann.  The second equation implies that $\momega$ belongs
to the cohomology class $\f\in H^1(M,\R)$ of $\mathbf{f}$. The
restriction $\cF\eqdef \bcF|_\cU$ is a regular codimension one
foliation which satisfies Theorems 1, 2 and 3 of \cite{g2} (which are
local in nature); those theorems give a complete characterization of
the intrinsic and extrinsic geometry of $\cF$. Using a certain
``improved parameterization'' of the four-form $F$ along the non-chiral locus $\cU$
(a parameterization which is ``adapted'' to the foliation $\cF$), a
lengthy computation shows that the results of \cite{g2} agree with
those of \cite{Tsimpis} along this locus; we refer the reader to
\cite{g2s} for details.

\section{Description of the singular foliation in the Morse case}
\label{sec:Morse}

Consider the case when the closed one-form $\momega\in \Omega^1(M)$ is
Morse. This case is generic in the sense that Morse one-forms
constitute a dense open subset of the set of all closed one-forms
belonging to the fixed cohomology class $\f$ --- hence a form
$\momega$ which satisfies equations \eqref{meq} can be replaced by a
Morse form by infinitesimally perturbing $b$.  Singular foliations
defined by Morse 1-forms were studied, for example, in \cite{Gelbukh3,
  Farber, Levitt3}.  Let $\Pi_f=\im (\per_\f)\subset \R$ be the period
group of the cohomology class $\f$ and $\rho(\f)=\rk \Pi_\f$ be its
irrationality rank.  A leaf $\cL$ of $\bcF$ is called {\em singular}
if it intersects $\cW$ and {\em regular} otherwise. Notice that $\cW$
is a finite set when $\momega$ is Morse. The study of Morse 1-forms is
a rich subject originating in Novikov theory, which is a
generalization of Morse theory from functions to forms. We refer the
reader to \cite{Farber} for an introduction to this subject.

\subsection{Types of singular points} 
\label{subsec:singtypes}
Let $\ind_p(\momega)$ denote the Morse index of a point $p\in \Sing
(\momega)=\cW$, i.e. the Morse index at $p$ of a Morse function $h_p\in
\cC^\infty(U_p,\R)$ such that $\dd h_p$ equals $\momega|_{U_p}$, where
$U_p$ is a sufficiently small vicinity of $p$. The Morse index does not
depend on the choice of $U_p$ and $h_p$. Let:
\beqa
&&\Sing_k(\momega)\eqdef \{p\in \cW|\ind_p(\momega)=k\}~~,~~k=1,\ldots, d\\
&&\Sigma_k(\momega)\eqdef \{p\in \cW|\ind_p(\momega)=k~\mathrm{or}~\ind_p(\momega)=d-k\}~~,~~k=1,\ldots,\left[\frac{d}{2}\right]~~. 
\eeqa
Thus $\Sigma_k(\momega)=\Sing_k(\momega)\cup\Sing_{n-k}(\momega)$ for
$k<\frac{d}{2}$ and $\Sigma_{d_0}(\momega)=\Sing_{d_0}(\momega)$ when $d=2d_0$
is even.
In a small enough vicinity of $p\in
\Sing_k(\momega)$ (which we can assume to equal $U_p$ by shrinking the
latter if necessary), the Morse lemma applied to $h_p$ implies that
there exists a local coordinate system $(x_1,\ldots,x_d)$ such that:
\beq
h_p=-\sum_{j=1}^k{x_j^2}+\sum_{j=k+1}^{d} x_j^2~~.\nn
\eeq

\paragraph{Definition.} The elements of $\Sigma_0(\momega)$ are called {\em centers}
while all other singularities of $\momega$ are called {\em saddle
  points}. The elements of $\Sigma_1(\momega)$ are called {\em strong
  saddle points}, while all other saddle points are called {\em weak}.

\paragraph{Remark.} Other names for the various types of singular points are in use 
in the Mathematics literature.

\subsection{Behavior of the singular leaves near singular points}

In a small enough vicinity of $p\in \Sing_k(\momega)$, the singular
leaf $\cL_p$ passing through $p$ is modeled by the locus $Q_k\subset
\R^n$ given by the equation $h_p=0$, where $p$ corresponds to the
origin of $\R^n$. One distinguishes the cases (see Tables
\ref{table:singtypes} and \ref{table:strongsaddles}):

\begin{itemize}
\itemsep 0.0em
\item $k\in \{0, n\}$, i.e. $p$ is a {\em center}. Then $\cL_p=\{p\}$ and
  the nearby leaves of $\cF_p$ are diffeomorphic to $S^{n-1}$.
\item $2\leq k\leq n-2$, i.e. $p$ is a {\em weak saddle point}. Then $Q_k$
  is diffeomorphic to a cone over $S^{k-1}\times S^{n-k-1}$ and
  $\R^n\setminus Q_k$ has two connected components while
  $Q_k\setminus\{p\}$ is connected. Removing $p$ does not {\em
    locally} disconnect $\cL_p$.
\item $k\in \{1,n-1\}$, i.e. $p$ is a {\em strong saddle point}. Then $Q_k$
  is diffeomorphic to a cone over $\{-1,1\}\times S^{n-2}$ and
  $\R^n\setminus Q_k$ has three connected components while
  $Q_k\setminus\{0\}$ has two components.  Removing $p$ {\em locally}
  disconnects $\cL_p$. 
\end{itemize}

\begin{table}[h!]
\centering
{\footnotesize 
\begin{tabular}{ | c | c| c| c| }
\hline
Name & Morse index & Local form of $\cL_p$ & Local form of regular leaves\\ 
\hline
Center & $0$ or $n$ & $\bullet=\{p\}$ &
\begin{minipage}{.2\textwidth}
\centering
\vspace{0.2em}
\includegraphics[scale=0.2]{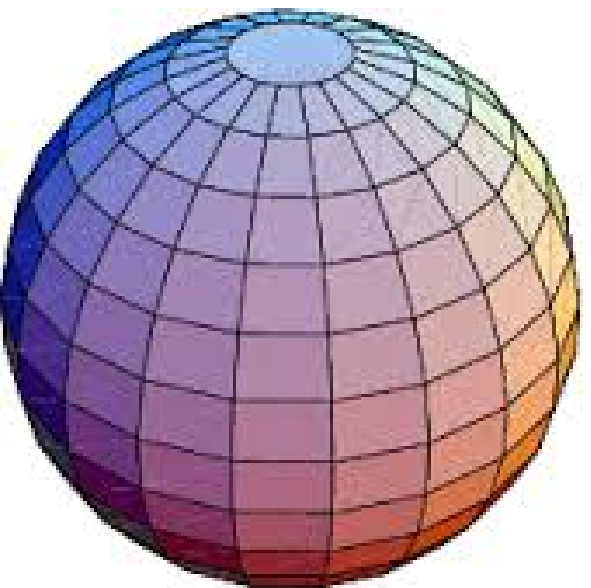}
\vspace{0.2em}
\end{minipage}\\ 
\hline
Weak saddle &  between $2$ and $n-2$ & 
\begin{minipage}{.2\textwidth}
\centering
\vspace{0.2em}
\includegraphics[scale=0.2]{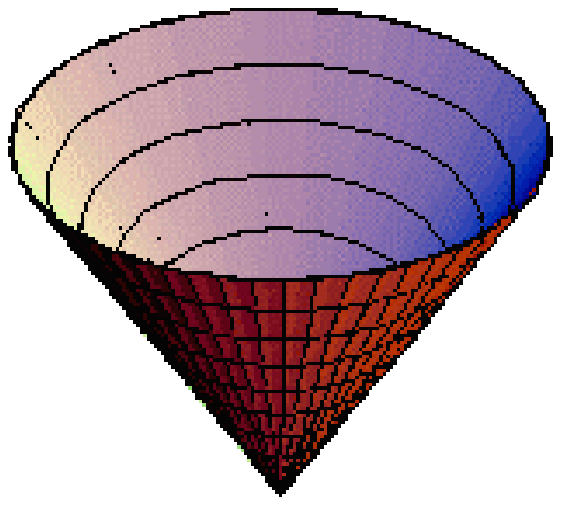}
\vspace{0.2em}
\end{minipage} 
& \begin{minipage}{.2\textwidth}
\centering
\vspace{0.2em}
\includegraphics[scale=0.2]{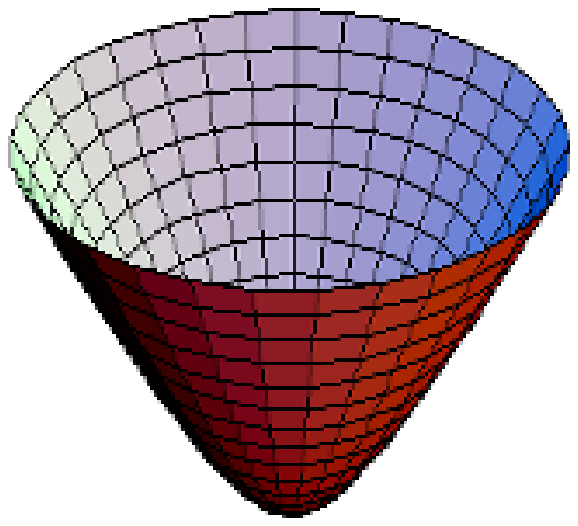}
\vspace{0.2em}
\end{minipage}\\ 
\hline
Strong saddle &  $1$ or $n-1$ &
\begin{minipage}{.2\textwidth}
\centering
\vspace{0.2em}
\includegraphics[scale=0.2]{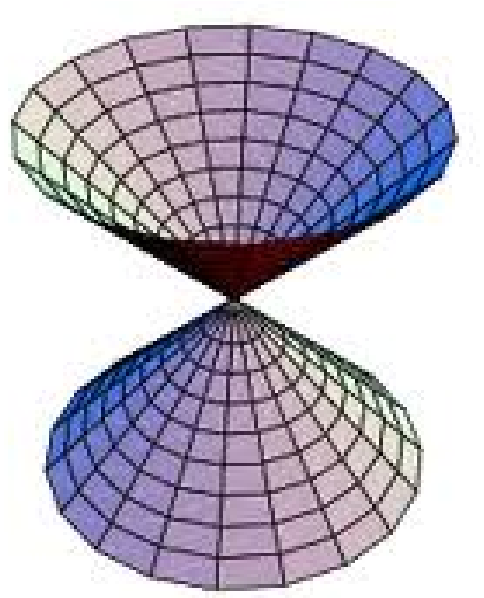}
\vspace{0.2em}
\end{minipage} 
& 
\begin{minipage}{.2\textwidth}
\centering
\vspace{0.2em}
\includegraphics[scale=0.2]{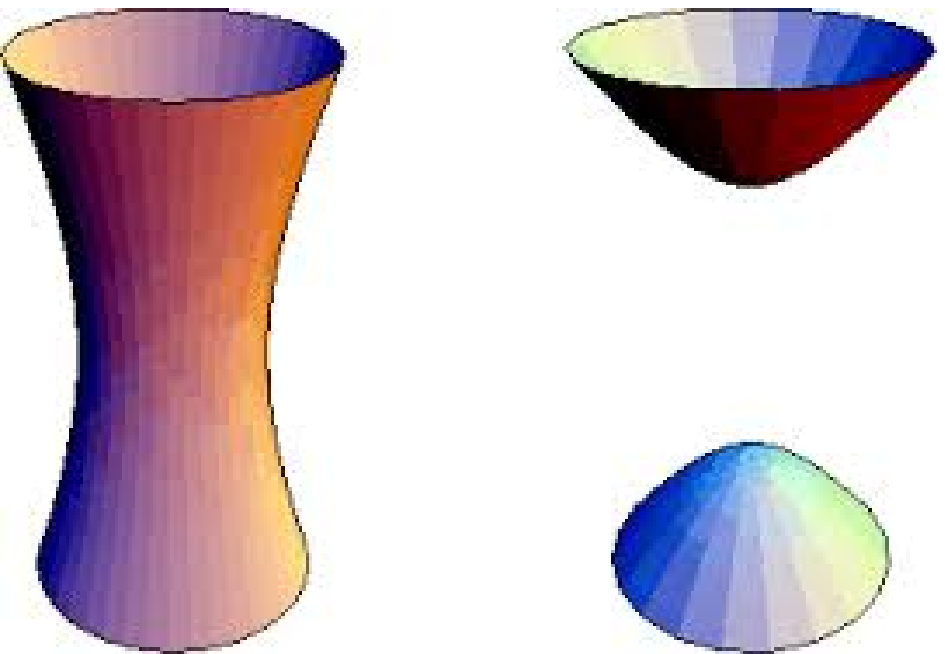}
\vspace{0.2em}
\end{minipage}\\ 
\hline
\end{tabular} }
\caption{Types of singular points $p$. The first and third figure on
  the right depict the case $d=3$ for centers and strong saddles,
  while the second figure attempts to depict the case $d>3$ for a weak
  saddle.}
\label{table:singtypes}
\end{table}

\begin{table}[h!]
\centering
{\footnotesize
\begin{tabular}{ | c | c | }
\hline
Singularity type & Example of global shape for $\cL_p$ \\ 
\hline
Splitting & 
\begin{minipage}{.3\textwidth}
\centering
\vspace{0.6em}
\includegraphics[scale=0.08, angle=90]{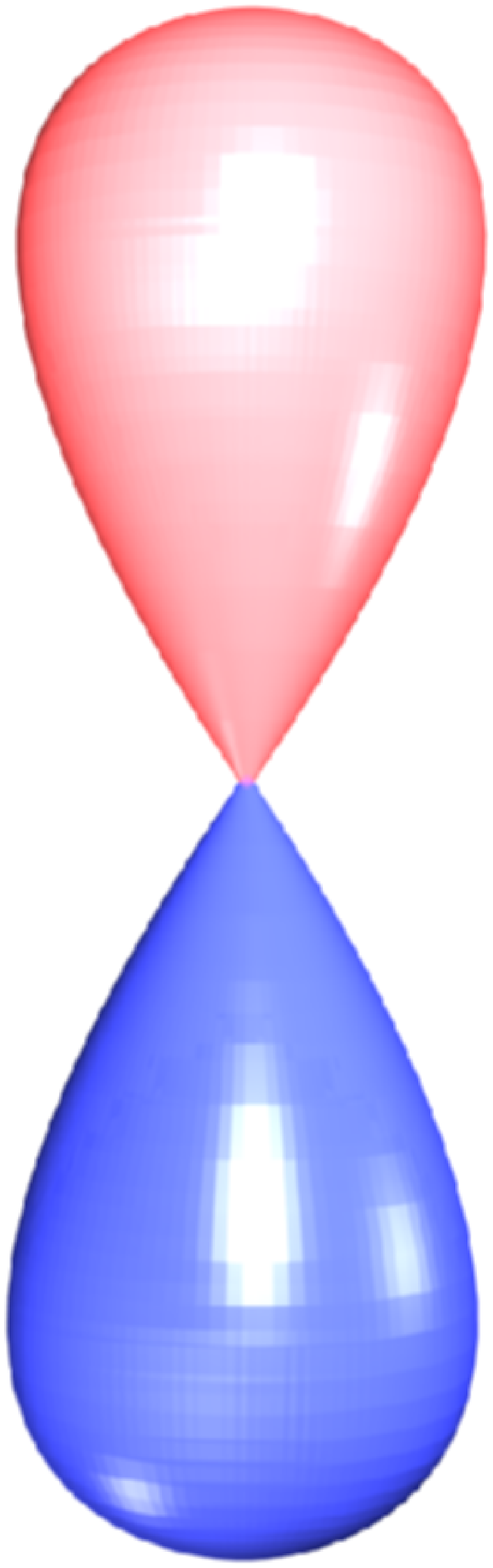}
\vspace{0.6em}
\end{minipage}\\ 
\hline
Non-splitting &  
\begin{minipage}{.3\textwidth}
\centering
\vspace{0.2em}
\includegraphics[scale=0.4]{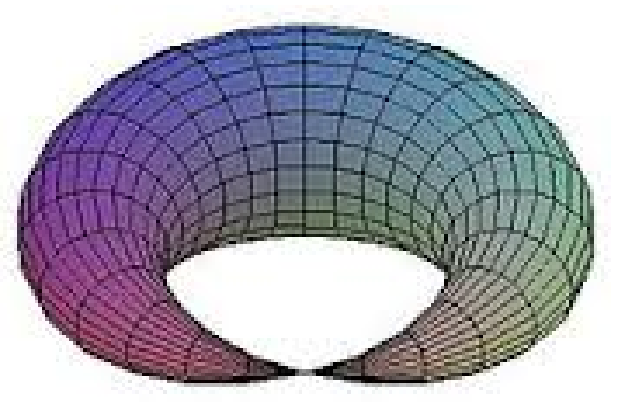}
\vspace{0.2em}
\end{minipage}\\ 
\hline
\end{tabular}}
\caption{Types of strong saddle points.}
\label{table:strongsaddles}
\end{table}

\subsection{Combinatorics of singular leaves}

\paragraph{Definition.} 
A singular leaf of $\bcF$ which is not a center is called a
{\em strong singular leaf} if it contains at least one strong saddle
point and a {\em weak singular leaf} otherwise.

A weak singular leaf is obtained by adjoining weak saddle
points to a single special leaf of $\cF$.  

\begin{figure}[h!]
\begin{center}
\includegraphics[scale=0.37]{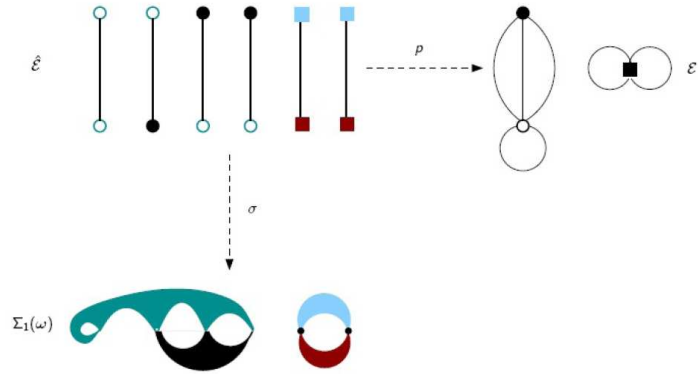}
\caption{Example of the graphs ${\hat \cE}$ and $\cE$ for a Morse form
  foliation $\bcF$ with two compact strong singular leaves.  
  The regular foliation $\cF$ of
  $M^\ast$ has four special leaves, depicted using four different colors, 
  each of which is compactifiable. At the bottom of the
  picture, we depict $\Sigma_1(\momega)$ as well as the schematic
  shape of the special leaves in the case $d=3$. The strong singular leaves
  of $\bcF$ correspond to the left and right parts of the
  figure at the bottom; each of them is a union of two special leaves
  of $\cF$ and of singular points. Each special leaf
  corresponds to a vertex of $\cE$.}
\label{fig:dimer}
\end{center}
\end{figure}

The situation is more complicated for strong singular leaves.
At each $p\in \Sigma_1(\momega)$, consider the strong singular leaf
$\cL$ passing through $p$. The intersection of $\cL\setminus\{p\}$
with a sufficiently small neighborhood of $p$ is a disconnected manifold
diffeomorphic to a union of two cones without apex, whose rays near $p$ determine a
connected cone $C_p\subset T_pM$ inside the tangent space to $M$ at
$p$ (see the last row of Table \ref{table:singtypes}). The set
$\dotC_p\eqdef C_p\setminus\{0_p\}$ (where $0_p$ is the zero vector of
$T_p M$) has two connected components, thus $\pi_0(\dotC_p)$ is a
two-element set. Hence the finite set:
\beq
{\hat \Sigma_1}(\momega)\eqdef \sqcup_{p\in \Sigma_1(M)} \pi_0(\dotC_p)
\eeq
 is a
double cover of $\Sigma_1(\momega)$ through the projection $\sigma$
that takes $\pi_0(\dot
C_p)$ to $\{p\}$. Consider the complete
unoriented graph having as vertices the elements of ${\hat
  \Sigma_1}(\momega)$.  This graph has a dimer covering given by the
collection of edges:
\beq
{\hat \cE}=\{\pi_0(\dotC_p)|p\in \Sigma_1(\momega)\}~~,
\eeq
which connect vertically the vertices lying above the same point of
$\Sigma_1(\momega)$ (see Figure \ref{fig:dimer}).  If $L$ is a special
leaf of $\cF$ and $p\in \Sigma_1(\momega)$ adjoins $L$, then
the connected components of the intersection of $L$ with a
sufficiently small vicinity of $p$ are locally approximated at $p$ by
one or two of the connected components of $\dotC_p$. The second case
occurs iff $p$ is a non-splitting strong saddle point (see Table
\ref{table:strongsaddles}).  Hence $L$ determines a subset ${\hat
  s}_1(L)$ of ${\hat \Sigma}_1(\momega)$ such that $\sigma({\hat
  s}_1(L))=s_1(L)$ and such that the fiber of ${\hat s}_1(L)$ above a
point $p\in s_1(L)$ has one element if $p$ is a splitting singularity
and two elements if $p$ is non-splitting.

 The graph $\cE$ has one vertex for each special leaf of $\cF$ which
adjoins some strong saddle point and an edge for each strong saddle
point. Notice that this edge is a loop when the strong saddle point
is a non-splitting singularity.

\noindent In our application, the set
$\Sing(\momega)=\cW=\cW^+\sqcup\cW^-$ consists of positive and
negative chirality points of $\xi$, which are the points where $b$
attains the values $b=\pm 1$.

\subsection{The foliation graph}

Define $C^\M$ to be the union of all compact leaves and $C^\m$ to be
the union of all non-compactifiable leaves of $\cF$ (see \cite{g2s}
for details).  Both $C^\M$ and $C^\m$ are open subsets of $M$ which
have a common topological small frontier\footnote{The {\em small frontier} of a set $A$ is the set $\fr(A)\eqdef A\setminus \Int A$.} 
$F$.  Each of the open sets $C^\M$ and $C^\m$ has a finite number of connected components, which
are called the {\em maximal} and {\em minimal} components of the set
$M\setminus F=C^\M\sqcup C^\m$ and we index them as $C^\M_j$ and
$C^\m_a$ such that:
\beq
C^\M=\sqcup_{j}C_j^\M~~,~~C^\m=\sqcup_{a}C_a^\m~~. \nn
\eeq
 Let: 
\beq
\bDelta\eqdef M\setminus C^\M=\overline{C^\m}=C^\m\sqcup F~~\nn
\eeq
be the union of all non-compact leaves and singularities. This subset
has a finite number of connected
components $\bDelta_s$.

\paragraph{Definition.} 
The {\em foliation graph} $\Gamma_\momega$ of $\momega$ is the unoriented
graph whose vertices are the connected components $\bDelta_s$ of
$\bDelta$ and whose edges are the maximal components $C_j^\M$. An edge
$C_j^\M$ is incident to a vertex $\bDelta_s$ iff a connected component
of $\fr C_j^\M$ is contained in $\bDelta_s$; it is a loop at $\bDelta_s$ iff~ 
$\fr C_j^\M$ is connected and contained in $\bDelta_s$. A vertex $\bDelta_s$ of
$\Gamma_\momega$ is called {\em exceptional} (or of {\em type II}) if it contains at least one
minimal component; otherwise, it is called {\em regular} (or of {\em type I}).
\begin{figure}[h!]
\begin{center}
\includegraphics[scale=0.33]{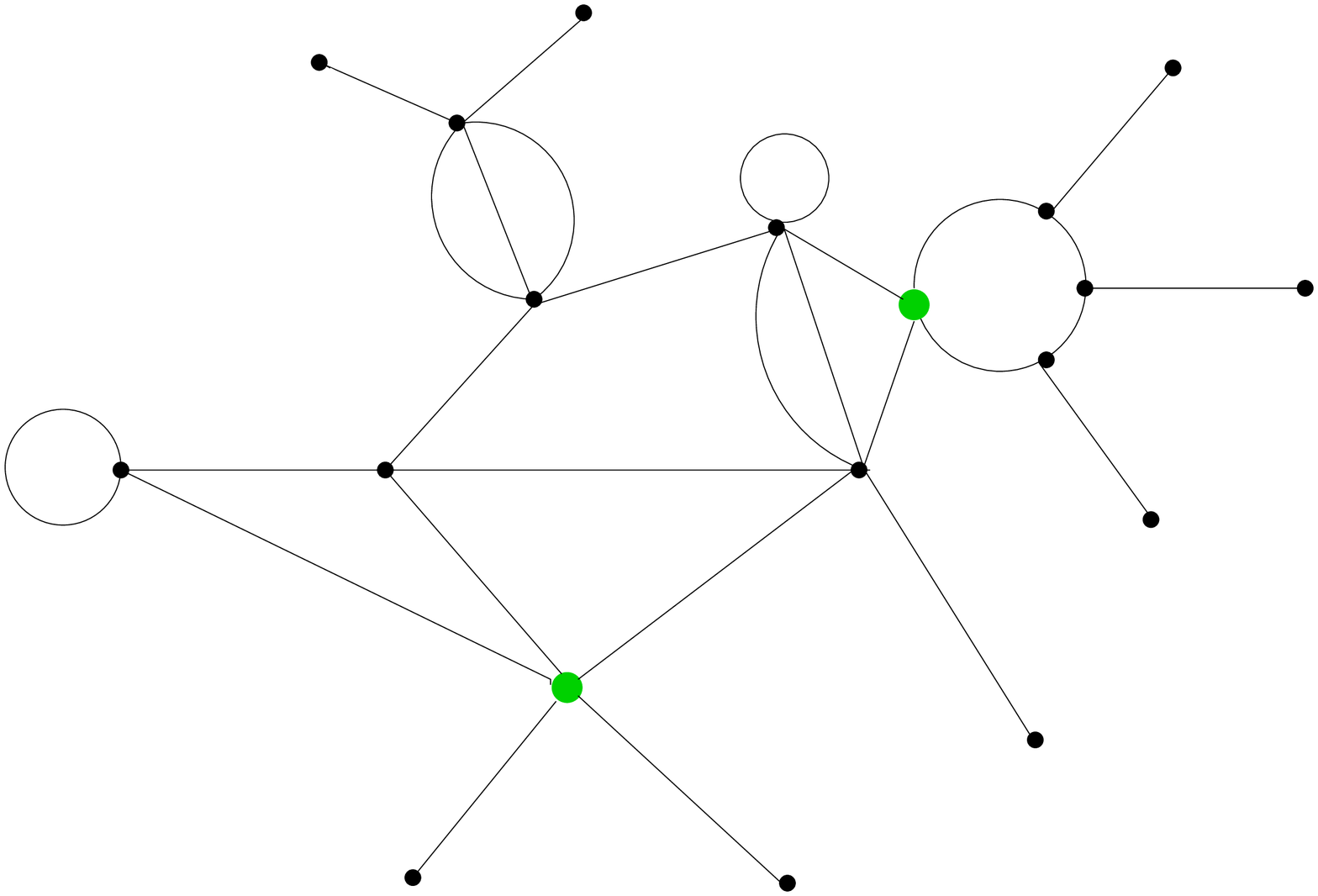}
\caption{An example of foliation graph. Regular (a.k.a type I)
  vertices are represented by black dots, while exceptional
  (a.k.a. type II) vertices are represented by green blobs. All
  terminal vertices are regular vertices and correspond to center
  singularities.}
\end{center}
\end{figure}

\begin{figure}[h!]
\begin{center}
\includegraphics[scale=0.3]{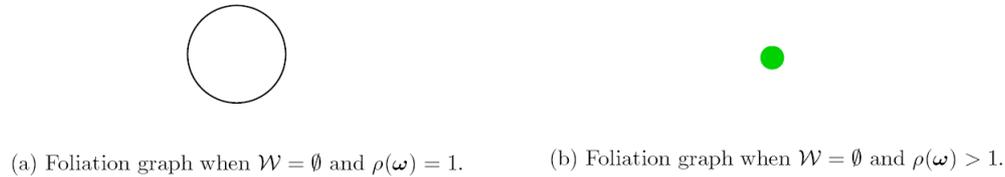}
\caption{Degenerate foliation graphs in the everywhere non-chiral case.}
\end{center}
\end{figure}
It is believed \cite{Levitt3} that the leaf space of $\bcF$ should be
described as a non-commutative space, the `commutative part' of which
is given by the foliation graph. A rigorous definition of the
$C^\ast$-algebra of singular foliations in the sense of Haefliger does
not appear to have been given in the Mathematics literature, so this
expectation should be taken with a grain of salt. Much more detail
about the topology of $\bcF$ can be found in \cite{g2s}. 

When $\xi$ is everywhere non-chiral (Case 3 of the topological no-go theorem,
i.e. $\cW=\emptyset$), the foliation graph is either a circle or a
single exceptional vertex (see Figure 3). It was shown in \cite{g2}
that, in this case, the exceptional vertex corresponds to a
non-commutative torus of dimension given by the projective
irrationality rank of $\momega$. Already in that case, one cannot
think of the generic compactification of this type (which corresponds
to a non-commutative leaf space) as a two-step reduction in the sense
of ``generalized Scherk-Schwarz compactifications with a twist'' (see
\cite{Vandoren}); this is doubly true when $\cW$ is a proper subset of
$M$.

\begin{acknowledgement}
E.M.B. acknowledges funding from the strategic grant POSDRU/159/1.5/S/
133255, Project ID 133255 (2014), co-financed by the European Social
Fund within the Sectorial Operational Program Human Resources
Development 2007--2013, while the work of C.I.L. is supported by the
research grant IBS-R003-G1.  This work was also financed by the
CNCS-UEFISCDI grants PN-II-ID-PCE 121/2011 and 50/2011, and by PN 09
370102.
\end{acknowledgement}

\end{document}